# Topological Semimetal Nanostructures: From Properties to Topotronics

An-Qi Wang,[1,2] Xing-Guo Ye,[1] Da-Peng Yu,[3] Zhi-Min Liao,[1,4*]

[1] State Key Laboratory for Mesoscopic Physics and Frontiers Science Center for Nano-optoelectronics, School of Physics, Peking University, Beijing 100871, China.

[2] Academy for Advanced Interdisciplinary Studies, Peking University, Beijing 100871, China.

[3] Shenzhen Institute for Quantum Science and Engineering and Department of Physics, Southern University of Science and Technology, Shenzhen 518055, China

[4] Collaborative Innovation Center of Quantum Matter, Peking University, Beijing 100871, China.

*E-mail: liaozm@pku.edu.cn

**ABSTRACT**: Characterized by bulk Dirac or Weyl cones and surface Fermi-arc states, topological semimetals have sparked enormous research interest in recent years. The nanostructures, with large surface-to-volume ratio and easy field-effect gating, provide ideal platforms to detect and manipulate the topological quantum states. Exotic physical properties originating from these topological states endow topological semimetals attractive for future topological electronics (topotronics). For example, the linear energy dispersion relation is promising for broadband infrared photodetectors, the spin-momentum locking nature of topological surface states is valuable for spintronics, and the topological superconductivity is highly desirable for fault-tolerant qubits. For real-life applications, topological semimetals in the form of nanostructures are necessary in terms of convenient fabrication and integration. Here, we review the recent progresses in topological semimetal nanostructures and start with the quantum transport properties. Then topological semimetal based electronic devices are introduced. Finally, we discuss several important aspects that should receive great effort in the future, including controllable synthesis, manipulation of quantum states, topological field effect transistors, spintronic applications, and topological quantum computation.

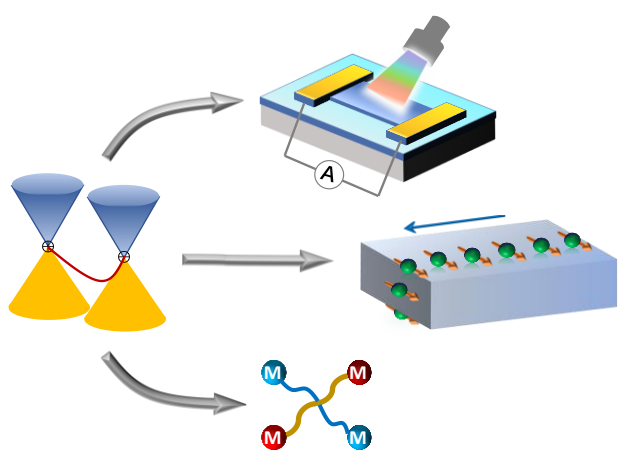

In condensed-matter physics, the discovery and classification of materials continue to attract great interest. The band theory of solids divides the common materials into three main types, metals, insulators and semiconductors, based on their electronic structures. Meanwhile, the various phases of materials can be classified using Landau's approach, which describes states in terms of fundamental symmetries and spontaneous symmetry breaking. Over the past four decades, the research of quantum

Hall effect has given rise to a totally different classification paradigm, based on the concept of topological phase.[1, 2] The emergent field of topological matter strongly motivates the development of topological materials.[3-8] The special band topology endows topological materials with specific electronic states, which are often topologically protected and immune to the environment perturbations, promising for low-dissipation and high-reliability electronic applications.

Among the various topological materials, topological semimetals have sparked intensive research interest due to bulk Dirac or Weyl fermions and nontrivial topological surface states.[6-8] The topological semimetals can be simply identified into Weyl semimetals and Dirac semimetals. Weyl (Dirac) semimetals are three-dimensional topological materials where valence band and conduction band touch at finite specific points in momentum space, named Weyl (Dirac) nodes. Generally, the Weyl (Dirac) semimetals could possess different dispersion powers ($k^n$) along different directions, resulting from their crystalline symmetry. Near the Weyl/Dirac node, the energy dispersion is linear along the primary rotation axis, while the dispersion along in-plane directions could be either linear ($n = 1$), quadratic ($n = 2$) or cubic ($n = 3$).[7, 9] A Weyl node with in-plane dispersion power $n$ corresponds to the degeneracy of $n$ conventional Weyl fermions (left-handed or right-handed) with same chirality.[10] By contrast, a Dirac node with $n$ corresponds to the degeneracy of $2n$ Weyl fermions, with half left-handed and half right-handed chirality.[11-13] The quadratic and cubic Weyl or Dirac semimetals are predicted to bring about exotic quantum physics, including non-Fermi liquid, quantum criticality and phase transition.[11, 14-16] In reality, the material realization of such quadratic and cubic semimetals remains scarce, and ever-identified topological semimetals mostly belong to the linear class, which harbors the typical Weyl or Dirac fermion in the low-energy limit. For simplicity, we mainly discuss the conventional linear Weyl and Dirac semimetals in this review.

For Weyl semimetals, the Weyl nodes always appear in pairs and inherit opposite chirality, such that the net chirality within the Brillouin zone is zero. Near each isolated Weyl node, the charges behave as relativistic Weyl fermions with linear energy dispersion in three dimensions. The chiral Weyl fermions lead to various exotic phenomena such as chiral anomaly,[17-23] anomalous Hall effect[24-27] and chiral magnetic effect.[28-31] The realization of Weyl semimetals generally requires the breaking of either inversion symmetry (IS) or time-reversal symmetry (TRS).[24] TRS-breaking systems can realize the minimal case of Weyl semimetal with only two Weyl points, by contrast, time-reversal invariant IS-breaking Weyl semimetals at least have four Weyl points in momentum space. Many IS-breaking Weyl semimetals have been theoretically proposed and then experimentally confirmed through angle-resolved photoemission spectroscopy (ARPES) and scanning tunneling microscopy (STM), such as the TaAs family,[32-40] layered transition-metal dichalcogenides $WTe_2$,[41, 42] $MoTe_2$,[43-47] and the alloy $Mo_xW_{1-x}Te_2$,[48-50] $TaIrTe_4$.[51, 52] Compared to the great number of IS-breaking Weyl semimetals, TRS-breaking Weyl semimetals still remain elusive. The TRS-breaking Weyl phase has been only predicted in a few magnetically ordered material candidates, including $Y_2Ir_2O_7$,[24] $GdPtBi$,[53] $HgCr_2Se_4$[27] and certain $Co_2$-based Heusler compounds.[54-57] Recently, a series of spectroscopic and transport experiments have confirmed the intrinsic magnetic Weyl phase in $Co_3Sn_2S_2$ and $Co_2MnGa$.[58-62] Magnetic Weyl semimetals have offered a platform to investigate the interplay between magnetism and topological phase, which can generate abundant exotic quantum states, ranging from quantum anomalous Hall effect (QAHE) to topological axion state.[7, 24, 58, 63-65] Aside from the magnetic Weyl semimetals, magnetic topological insulators have attracted broad attention lately. The magnetic topological insulator (TI) phase has been identified in some ferromagnetic candidates, such as $EuSn_2As_2$, $MnBi_2Te_4$ and $MnBi_4Te_7$, through the observations of quantized Hall conductivity plateau,[66, 67] QAHE[68] and ARPES results.[69-73] It's proposed that the ferromagnetic phase of $MnBi_2Te_4$-family materials could also lead to an ideal magnetic Weyl semimetal.[74-77]

As mentioned above, Weyl semimetals are formed under the breaking of either TRS or IS. Otherwise, a Dirac semimetal would come into being if two such symmetries are simultaneously preserved. Near the nodal points, the low-energy quasiparticle excitations in Dirac semimetals behave as Dirac fermions, which are degenerate and can be viewed as a superposition of Weyl fermions with opposite chirality. Generally, the overlap of two Weyl nodes would induce band hybridization and gap opening.[10] So additional crystal symmetry, such as $C_4$ rotational symmetry, is required for the stability of a three-dimensional (3D) Dirac semimetal.[78] Upon breaking either time-reversal symmetry or inversion symmetry, the Dirac semimetal would transform into the Weyl semimetal by means of splitting each Dirac point into two Weyl nodes. In addition, Dirac semimetals can also form other nontrivial phases including topological insulator and quantum spin Hall insulators.[79] Following the theoretical predictions that some materials are Dirac semimetals, including $Na_3Bi$ and $Cd_3As_2$,[78, 79] a series of ARPES experiments are carried out to confirm the presence of the special band structure.[80-86] Meanwhile, STM measurements are also used to demonstrate the Dirac semimetal phase.[87]

Besides the bulk relativistic fermions, the topological semimetals also possess topological surface states with nontrivial electronic properties. In a Weyl semimetal, the surface states are predicted to exist in the form of open Fermi arcs. Each Fermi arc originates from and terminates at the surface projection of bulk Weyl nodes of opposite chirality, which has been experimentally identified by ARPES in many materials.[33-35, 80-86, 88, 89] Unlike closed Fermi surface in conventional materials, the open Fermi arcs and their interplay with bulk Weyl nodes can give rise to various exotic transport phenomena, including π Aharonov–Bohm (AB) effect,[90, 91] Weyl-orbit related quantum oscillations,[92-96] quantum Hall effect,[97-101] Fano interference,[102] unusual quasiparticle interference,[103-106] as well as the emergent Majorana zero modes when coupled with s-wave superconductivity.[107-109] Contrary to the case of Weyl semimetals, the surface states of Dirac semimetals seem more complicated and still remain under debate.[7, 110] Considering the bulk Dirac nodes can be viewed as two degenerate Weyl nodes, it is natural to expect the surface states in a Dirac semimetal are two copies of chiral Fermi arcs.[78, 79] The signatures of Fermi arcs have been reported in Dirac semimetal $Na_3Bi$ and $Cd_3As_2$, through the ARPES[86] and magneto-transport measurements,[92, 95] respectively. However, some recent studies demonstrate that, unlike the case of Weyl semimetals, the Fermi arcs in a Dirac semimetal are not topologically protected and fragile to environmental perturbations.[110, 111] The open Fermi arcs would possibly deform into a closed Fermi contour as varying the chemical potential.[110]

With the recent advances of "higher-order topological phase" theory, the lower-dimensional boundary states, such as one-dimensional (1D) hinge states and zero-dimensional (0D) corner states are predicted in topological materials.[112-120] According to the theory, a $D$-dimensional $k$th order topological phase holds gapless boundary states with ($D$-$k$)-dimension. Various examples of higher-order topological phases, protected by spatiotemporal symmetries, have been discussed, including 3D insulators with chiral (helical) hinge states[113, 114, 119] and two-dimensional (2D) second-order insulators with corner states.[112, 113] Beyond theoretical predictions, the higher-order topological phase has been realized in bismuth, and its hinge states are experimentally confirmed.[120] Additionally, there are a series of acoustic and photonic realizations of second-order topological insulator phase with corner states.[121-127] For a topological semimetal, besides the bulk Weyl/Dirac cones and surface Fermi arcs, it may support 1D hinge states and 0D corner states in its higher-order topological phase (Figure 1a-c).[115, 118, 128] Wieder *et al.* proposed the notion of higher-order Fermi arcs on 1D hinges, which connect the surface projections of bulk Dirac nodes.[129] Experimentally, Li *et al.* have recently identified the existence of 1D hinge states in Dirac semimetal $Cd_3As_2$ through the Josephson interferometry measurements.[130]

The inherent electronic properties render topological semimetals potential for future electronic applications. For example, the gapless band structure is promising for broadband photodetectors,[6-8] the spin momentum locking of Fermi arcs for spintronics,[86, 90, 91] and the helical nature of surface electrons for topological qubits.[107-109] For nanoscale device applications, the topological semimetals in the form of nanostructures are necessary. Figure 1d,e show two typical nanostructures, nanowires and nanoplates, respectively. From the perspective of fundamental physics, topological nanostructure is helpful to reveal special topology of exotic surface states, in which the influence of undesirable bulk shunt is greatly reduced.[6, 8] Besides, the nanostructure can ensure effective field-effect gating to tune the Fermi level and realize the manipulation of quantum states. Moreover, the nanostructure provides an excellent platform to study the exotic properties arising from quantum confinement in topological semimetals.[8] Considering the potential device application and underlying fundamental physics, research of topological semimetal nanostructures is promising and attractive. In this review, we mainly focus on recent progresses about the physical properties and device applications of topological semimetal nanostructures. We start with the transport properties from bulk Weyl/Dirac fermions. Then we introduce the quantum phenomena from topological surface states. After that, we summarize the recent advances in semimetal-based electronic devices, including the ultrafast broadband photodetectors, spin field-effect transistors and superconducting devices for topological qubits. Finally, we give a concise conclusion and look forward the future.

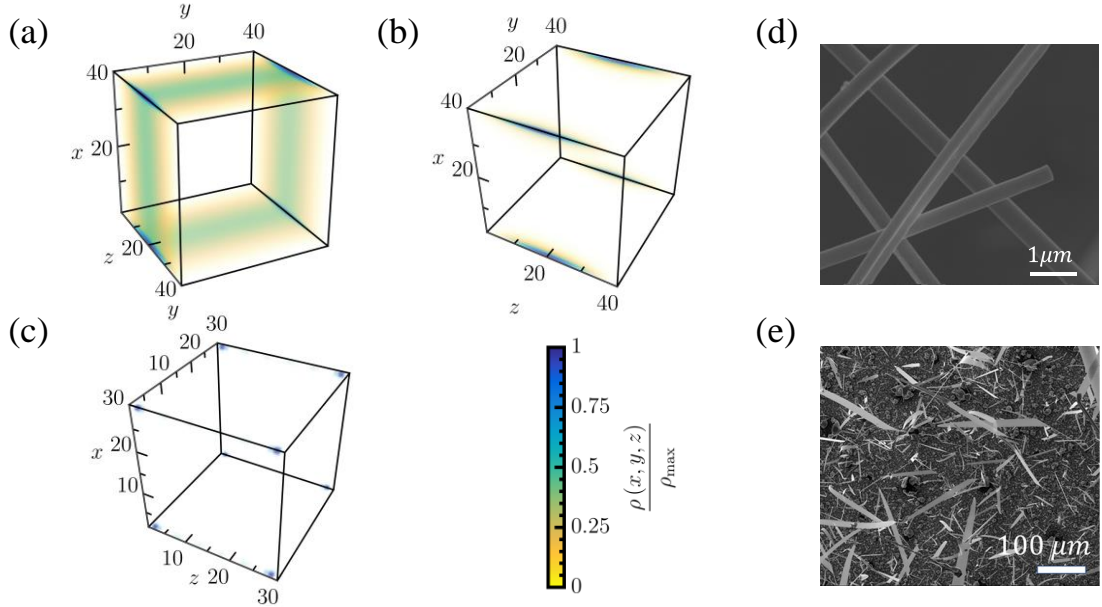

**Figure 1. Topological boundary states and nanostructures.** (a) 2D Fermi arc surface states of a first-order, (b) 1D hinge states of a second-order, and (c) 0D corner states of a third-order topological Dirac semimetal. (d) The scanning electron microscope (SEM) image of the $Cd_3As_2$ nanowires, which show a large surface-to-volume ratio. (e) The SEM image of the $Cd_3As_2$ nanoplates. Panels a-c reprinted with permission from ref 128. Copyright 2019 American Physical Society.

## QUANTUM TRANSPORT FROM THE BULK STATES

Abundant physical phenomena due to the exotic topology of bulk Weyl/Dirac fermions have been revealed through transport measurements. Chiral anomaly, characterized by charge pumping between opposite Weyl nodes, is regarded as one of the most distinctive properties in topological semimetals.[19,131-139] The negative MR is commonly used as an evidence of chiral anomaly in topological semimetals including $Bi_{1-x}Sb_x$,[134] $TaAs$,[135] $NbP$,[136] $Na_3Bi$,[137] $ZrTe_5$[138] and $Cd_3As_2$.[131] The Shubnikov-de Haas (SdH) oscillations,[140] with a nontrivial Berry phase $\pi$,[101] have been used as one of the transport evidence to confirm the nontrivial phase in topological semimetals, similar to the cases in graphene[141] or topological insulators.[142] Due to the spin-orbit interactions, the weak antilocalization (WAL) effect is expected to be observed within low magnetic fields at low temperatures in topological semimetals, but strong intervalley scattering can suppress WAL and lead to weak localization (WL).[143] Meanwhile, a parabolic MR has been observed when a low perpendicular magnetic field is applied in topological semimetals.[144] The $B^2$ dependent MR has been considered to originate from the classical cyclotron orbit of electrons driven by Lorentz force, which can be used to estimate the carrier mobility.[139] Additionally, a large linear MR has been found at high fields, usually accompanying with notable SdH oscillations.[145, 146] Such a positive linear MR demonstrates giant and non-saturating characteristics, which is totally different from conventional parabolic MR.[6] Lately, the nonlinear Hall effect, originating from Berry-curvature dipole, has also been observed in topological semimetals.[147-149] Below, we will expand on the chiral anomaly, SdH oscillations with Berry phase π, large linear MR and the nonlinear Hall effect from Berry-curvature dipole.

**Chiral Anomaly.** In the presence of a magnetic field and an electric field which are parallel with each other, there would be an imbalance of chemical potential between two Weyl nodes with opposite chirality, rendering a charge pumping from one Weyl node to the other with opposite chirality. And the chiral charge near a single Weyl node will satisfy $\frac{\partial n_{R/L}^{3D}}{\partial t} = \pm \frac{e^2}{h^2} \boldsymbol{E} \cdot \boldsymbol{B}$,[7] where $n_{R/L}^{3D}$ represents the chiral charge density, $\pm$ denotes the right- (R) and left-handed (L) chirality, $\boldsymbol{E}$ and $\boldsymbol{B}$ are the applied electric field and magnetic field, respectively. Obviously, the chiral charge at single Weyl node is not conserved, which is so-called chiral anomaly (Figure 2). As for Dirac semimetals, when magnetic fields are applied, the overlapping Weyl nodes will separate with each other along the magnetic field direction in momentum space, and the distance is proportional to the field strength (Figure 2a).[150] Thus, chiral anomaly can also emerge. The chiral anomaly effect can lead to abundant physical phenomena, including negative MR, anomalous thermoelectric effect,[132] nonlocal valley transport,[133] as well as magneto-optical Kerr effect.[133]

The negative MR has been observed in lots of topological semimetals,[131, 133, 138, 139, 151] which is attributed to chiral anomaly. When the $\boldsymbol{E} \cdot \boldsymbol{B}$ term exists, Weyl nodes with opposite chirality will have different chemical potentials ($\mu^R \neq \mu^L$) and charge pumping emerges, rendering a chiral current (Figure 2b). The chiral current can give positive contributions to the conductivity, leading to the negative MR (Figure 2c). It is worth noting that the negative MR can only be found in systems with low carrier density.[131] High carrier density will cause the Fermi energy above the Lifshitz point, where the Weyl nodes would be overwhelmed and thus no chiral anomaly effect observed.[78] As depicted in Figure 2d, the amplitude of negative MR decreases with the increase of gate voltage. This may also explain why the negative MR is usually absent in bulk materials. With increasing temperatures, the negative MR is still robust (Figure 2c), ruling out the influence of WL effect, in which negative MR only survives at low temperatures.[139] Angle-dependent measurements are also carried out to make sure the negative MR indeed originates from chiral anomaly.[131] When the magnetic field is tilted away from the electric field direction, the magnitude of the negative MR gradually decreases and finally disappears, which can be explained by the $\boldsymbol{E} \cdot \boldsymbol{B}$ term. In addition, the negative MR is also found in Cd$_3$As$_2$ nanoplates with thickness about several hundred nanometers.[131, 133]

Recently it is proposed that some other mechanisms can lead to negative MR in semimetal systems. For example, current jetting effect as reported in TaAs family.[151, 152] Besides, in the ultra-quantum limit,[18, 19, 153] finite ionic impurity can also induce negative MR in three-dimensional metal according to theoretical proposals.[154, 155] These mechanisms must be carefully ruled out before confirming the chiral anomaly as the origin of negative MR.

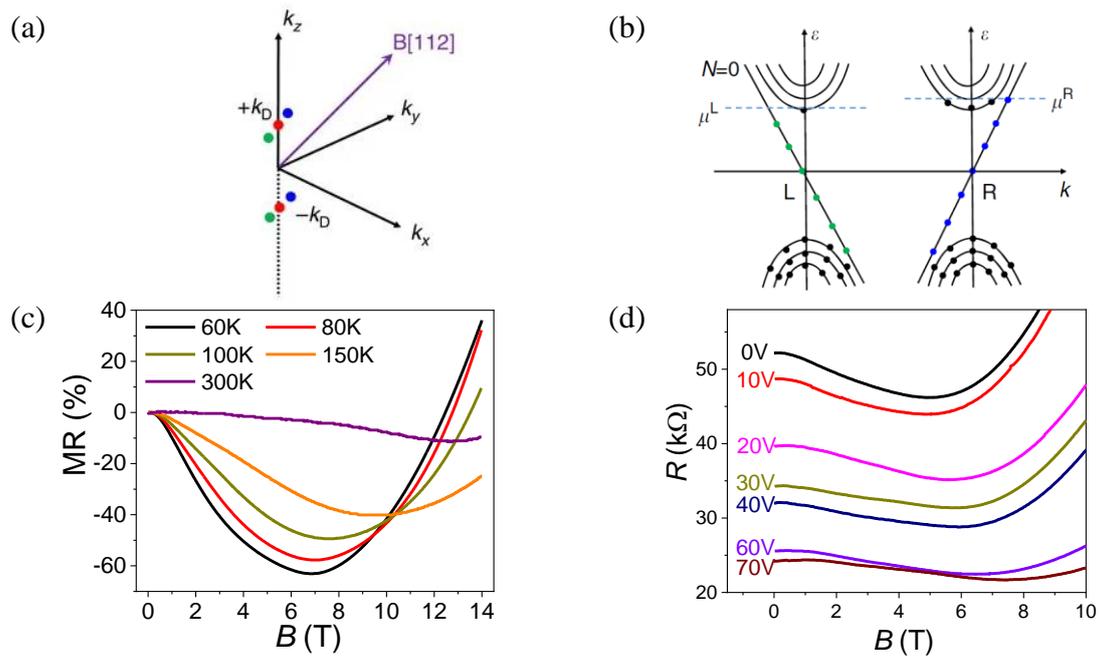

**Figure 2. Chiral anomaly induced the negative MR in a Cd$_3$As$_2$ nanowire.** (a) The diagram of a Dirac point splitting into a pair of Weyl points along the direction of external magnetic field. The red points denote the Dirac points, while the green and blue points denote the left-handed and right-handed Weyl points, respectively. (b) The diagram of chiral anomaly in Weyl semimetals. L and R represent the left-handed and right-handed chirality in the N = 0 Landau level, respectively. (c) The negative MR under a parallel magnetic field from 60 to 300 K. (d) The gate voltage modulated MR at 60 K. Panels a-d reprinted with permission under a Creative Commons CC BY License from ref 131. Copyright 2015 Springer Nature.

Besides the charge transport, heat transport can provide rich information about Weyl fermions. Thermoelectric effects have been predicted in the Weyl and Dirac semimetals.[156, 157] When the applied magnetic field is parallel to the temperature gradient, the chiral anomaly effect will introduce a parabolic field-dependent contribution to the longitudinal thermal conductivity.[156] Such thermal response of Weyl fermions has been experimentally observed in Dirac semimetal $Cd_3As_2$ nanoplate.[132] The thermal-power shows a suppressed quadratic dependence on the parallel field and reverses its sign with further increasing the field value. Such field-dependent thermal signal can be well explained with Motte relation,[158] combined with the chiral anomaly-induced magnetoconductivity and conventional Drude conductivity.[132] When the applied magnetic field deviates from the temperature gradient or the temperature is increased, the quadratic suppression of thermal response will be weakened, indicating the close relation with the chiral anomaly-induced magnetoconductivity. Besides the chiral anomaly, recent studies show that the chiral zero sound, arising from chiral magnetic effect of Weyl fermions, can also cause an anomalous thermoelectric effect in the Weyl semimetals.[159] The chiral zero sound describes a collective bosonic excitation, which can be viewed as an efficient heat carrier and leads to large ratio of heat over charge conduction. Since the velocity of chiral zero sound is inversely proportional to the electronic density of states at the Fermi level, strong magnetic quantum oscillations of thermal conductivity will take place along the field direction, akin to the conventional quantum oscillations of electric conductivity. More recently, Xiang *et al.* have reported the experimental observation of giant magnetic quantum oscillations of thermal conductivity in Weyl semimetal TaAs and proposed the chiral zero sound as the most likely cause.[160] The thermoelectric evidence of chiral zero sound hitherto remains elusive, calling for further experimental investigation on topological semimetals.

**SdH Oscillations with Berry Phase $\pi$.** SdH oscillation originates from the Landau quantization of electron orbits under high magnetic fields. The oscillation can be described by Lifshitz-Kosevich formula $\cos[2\pi(\frac{F}{B} + \phi)]$, where B is the magnitude of magnetic field, F is the oscillation frequency, and $\phi$ is the phase factor correlated with the Fermi surface topology.[140] It's widely believed that an energy band with linear dispersion would introduce a $\pi$ Berry phase,[161, 162] leading to $\phi=0$ and $\pm 1/8$ (+ for hole,− for electron carrier) in 2D and 3D system, respectively. In this regime, 3D topological semimetals are expected to harbor a phase $\phi = \pm 1/8$ in the SdH oscillations. The quantity $\phi$ can be experimentally accessed by analyzing the Landau fan diagram of SdH oscillations. A crucial issue here is that whether integer or half integer Landau indices correspond to the resistivity peaks in SdH oscillations. A recent theoretical work on Weyl and Dirac semimetals demonstrates that, the resistivity peaks appear at the Landau band edges and should be assigned integer indices.[140] Moreover, the phase factor $\phi$ of Dirac semimetals or Weyl semimetals with time-reversal symmetry is predicted to take the value of $\pm 1/8$ or $\pm 5/8$, which is consistent with the phase shift observed in previous experiments.[146, 163-165] In contrast, for Weyl semimetals with broken time-reversal symmetry, the phase factor undergoes a non-monotonic evolution with varying Fermi energy and approaches a wide range between $\pm 7/8$ and $\pm 9/8$ near the Lifshitz point.[140]

In addition, the SdH oscillations can also be used to analyze the geometry of Fermi surface of topological semimetals, like $Cd_3As_2$,[165] through the angle-dependent oscillation frequency. Most interestingly, the evolution of Berry phase in certain processes may be obtained by analyzing SdH oscillations,[164, 166] which can be used as an effective detection method to reveal the possible topological phase transition.

**Large Linear MR.** A giant positive linear MR has been observed in topological semimetals under high magnetic field.[135, 145, 163, 167] Many possible physical origins about the linear MR have been proposed. The Abrikosov theory,[168] known as the quantum linear MR model, shows that a non-saturating linear MR can occur in 3D gapless semiconductors with the linear dispersion when all electrons are filled into the first Landau level (LL), that is, the quantum limit.[6] However, a linear MR has been observed in some experiments within low fields where not all electrons are filled into the first LL.[144] Another possible theory is proposed by Parish and Littlewood,[169] arguing that the linear MR possibly originates from large mobility fluctuations induced by disorder effects. Beyond the two theories, people also explain the giant non-saturating MR in the context of electron-hole compensation[170] and field-induced relative shifting between Weyl-Fermi surfaces.[144] In fact, the linear MR has been observed in various systems including not only Dirac and Weyl semimetals, but also bismuth thin films,[171] InSb,[172] graphene[173] and topological insulators[174, 175] experimentally. There exist many possible theories to explain the linear MR in different systems.[144] Such a wide range of materials share similar behaviors, indicating that behind the linear behavior, there exists a more general physical origin.

**Nonlinear Hall effect from Berry-curvature dipole.** Upon breaking inversion symmetry, the gap opening at the Dirac points would lead to large Berry curvature near the gap edge.[162, 176] If the positive and negative Berry curvatures are segregated in $k$ space, a dipole moment comes into being (Figure 3). Such a Berry-curvature dipole would give rise to an unusual electrical effect, that is second-order nonlinear Hall effect.[177] The realization of conventional Hall effect requires breaking of time-reversal symmetry, usually in magnets or applying magnetic fields (Figure 3a,b). Contrary to the conventional Hall regime, the presence of nonlinear Hall effect needn't to break the time-reversal symmetry (Figure 3c,d),[177-179] which has already attracted huge research interest.[180-189] The second-order nonlinear Hall effect is characterized by a quadratic, rather than linear, current-voltage characteristic. Phenomenologically, the effect can be simply described as $V^{NLHE} \propto E^2$, where the $E$ is the external electric field along the direction of dipole moment, and $V^{NLHE}$ is the transverse Hall voltage generated by the electric field. An a.c. electric signal with frequency ω will generate a transverse Hall voltage with double frequency 2ω, which can be experimentally detected through lock-in techniques. Moreover, in the nonlinear Hall regime, the transverse Hall voltage is much larger than the longitudinal response, resulting in a Hall angle of nearly 90 degree. Recently, Ma *et al.* have reported the observation of nonlinear

Hall effect in bilayers of non-magnetic material WTe$_2$ under time-reversal-symmetric conditions.[147] Much larger than the longitudinal voltage, the transverse Hall voltage exhibits a quadratic current-voltage characteristic, together with the double frequency 2ω, revealing the nonlinear Hall nature (Figure 3e). Shan and Dzsaber *et al.* have also observed the nonlinear Hall signals induced by large Berry-curvature dipole in topological semimetal T$_d$-WTe$_2$ and Ce$_3$Bi$_4$Pd$_3$.[148, 149] The observation of nonlinear Hall effect gives a direction for exploring Berry curvature physics in non-magnetic materials and topological semimetals.

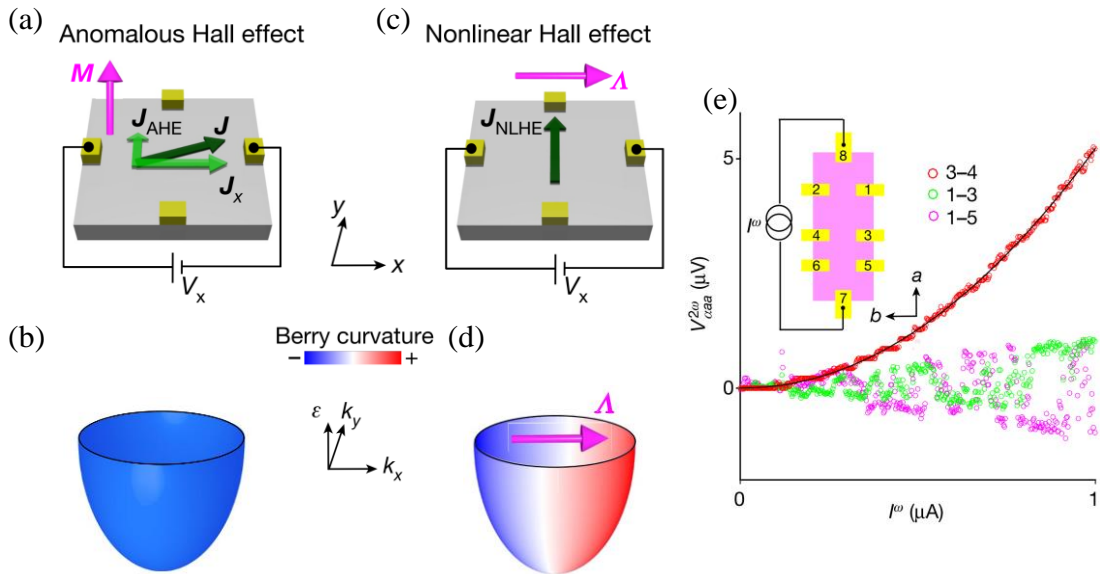

**Figure 3**. **Nonlinear Hall effect from the Berry-curvature dipole.** (a) Schematic of the anomalous Hall effect in a magnetic metal. ***J*$_x$** and ***J*$_{AHE}$** are the bias current and anomalous Hall current, respectively. ***J* = *J*$_x$ + *J*$_{AHE}$** is the total current. ***M*** denotes the magnetization. (b) Distribution of the Berry curvature of a magnetic metal. (c) Schematic of the nonlinear Hall effect. ***Λ*** represents the Berry-curvature dipole. When applying a bias current parallel to ***Λ***, a nonlinear Hall current ***J*$_{NLHE}$** is established in the transverse direction. (d) Distribution of the Berry curvature in non-magnetic, inversion-

symmetry-breaking quantum materials with non-zero $\Lambda$. $\Lambda$ arises from the segregation of positive and negative Berry curvature in $k$ space, which doesn't need to break the time-reversal symmetry. (e) In the presence of an a.c. current with frequency ω, the nonlinear voltages $V^{2\omega}$ along the longitudinal and transverse directions are detected. Panels a-e reprinted by permission from Springer Nature Customer Service Centre GmbH: Springer Nature ref 147, copyright 2018.

## QUANTUM TRANSPORT FROM THE SURFACE FERMI ARCS

On the surface of a Weyl semimetal, the surface state exists in the form of open Fermi arcs, which connects the surface projection points of bulk Weyl nodes, instead of closed Fermi loop[19, 24, 190, 191]. The Fermi arc can be understood as the edge state of a series of 2D topological insulators within two opposite-chirality Weyl nodes.[6] Alternatively, the Fermi arc can also be described in the regime of helicoidal structures and non-compact Riemann surfaces, where the bulk nodal points project to the branch points of the helicoids whose equal-energy contours are Fermi arcs.[192] But for Dirac semimetals, the Fermi arcs are fragile to environmental perturbance according to recent theoretical works,[110, 192-195] which may explain the Fermi arc is not commonly experimentally observed in Dirac semimetals. But notably, when breaking time-reversal symmetry (such as applying a magnetic field), Dirac semimetal can transform into Weyl semimetal and thereby inheriting the Fermi arc surface states.

Prior to bulk materials, the nanostructures usually inherit a higher crystal quality and has a much larger surface-to-volume ratio, providing a platform to interpret the Fermi arc related transport. In a topological Weyl/Dirac semimetal nanowire, the energy spectrum of the Fermi arc states becomes a series of quantized sub-bands due to the spatial confinement.[196] The separation between the discrete sub-bands decrease with the growth of the wire diameter, and these sub-bands eventually evolve to standard Fermi arc states that connect the bulk Fermi surfaces. In contrast to the flat Fermi arc bands in the semi-infinite slab, the energy band of nanowire has a weak parabolic momentum dependence along the direction of connecting the Weyl nodes, rendering the corresponding Fermi arc transport may be dissipative.[196] In a nanowire, the electric charge and current density distribution is spatially nonuniform since the rather large amount of charges is accumulated at the surface.[196] The contribution ratio of bulk and surface states can be modulated by tuning the Fermi level, where the lower Fermi level facilitates the dominance of Fermi arc contribution. In the regime of broken time-reversal symmetry, Weyl semimetal nanowires can support a nontrivial magnetization current mainly at the circumferential surface even when the external magnetic field is absent.[196] Besides the magnetization current, transport properties of Weyl semimetal nanowires are theoretically investigated, mainly focusing on the conductance calculations in different regimes.[107, 197, 198] For example, the Fermi arc states of Weyl semimetal nanowire are predicted to give rise to quantized conductance step with chemical potential.[198] In the presence of magnetic flux threading the nanowire, there will emerge conductance quanta with characteristic interference oscillations.

 The quantum transport properties of the Fermi-arc surface states have been revealed. Here we would like to expand on the topological surface states transport properties including $\pi$ AB effect,[90, 91] Fano effect,[102] and quantum Hall effect.[97-101]

**Aharonov-Bohm effect**. AB oscillations of a core-shell nanostructure, such as nanowire and nanoribbon, has become an effective way to prove the existence of surface states.[90, 199-207] Because in such core-shell structure, only the outmost surface states contribute to the effective AB phase, while the bulk phase from different cross-sections would be de-coherent, causing no AB effect observed.[208]

In a nanowire (or nanoribbon) system, the surface energy band would evolve into a series of discrete sub-bands, arising from the quantum confinement and quantized wave vector $k_\perp$ along the circumferential direction. In the presence of an axial magnetic field, the electrons would accumulate an additional AB phase of $2\pi\Phi/\Phi_0$ (where $\Phi$ is the magnetic flux penetrating the cross-section, and $\Phi_0 = h/e$ is the quantum flux, h is Plank's constant and e is electron charge).[207] Given the initial phase randomness of bulk states from different cross-sections, the overall bulk phase interference is destructive, and therefore AB effect is believed to only originate from the surface states. Additionally, if the surface states harbor a spin-helical nature, an extra Berry phase of π needs to be considered when the carrier travels along the whole circumference.[3, 4, 162, 209, 210] Taking both AB phase and Berry phase π into consideration, the quantized surface sub-band in a TI nanostructure can be simply expressed as[91, 199, 201, 207, 211]

$$E = \pm h v_F \sqrt{\frac{k^2}{4\pi^2} + \left(\frac{m+\frac{1}{2}-\frac{\Phi}{\Phi_0}}{C}\right)^2},$$

where $v_F$ is the Fermi velocity, k is the momentum vector along the axis, C is the circumference, and $m = 0, \pm1, \pm2...$ is the angular momentum quantum number. The term 1/2 represents the π Berry phase. Figure 4a depicts the evolution of surface sub-bands with magnetic fields. When $\Phi = m\Phi_0$, the sub-band spectrum is gapped. When $\Phi$ is an odd multiply of $\Phi = (m + 1/2)\Phi_0$, 1D linear gapless modes emerge. Notably the linear bands are nondegenerate while the parabolic bands are doubly degenerate. According to the structure of sub-bands, conductance valley would appear at $\Phi = m\Phi_0$ (π-AB effect), when the Fermi level is located at the Dirac point (blue dashed line). As tuning the Fermi level away from the Dirac point (red dashed line), conductance valley would appear at $\Phi = (m + 1/2)\Phi_0$ (AB effect). The transition from π-AB effect to AB effect of $Cd_3As_2$ nanowire is observed,[91] as shown in Figure 4b. Both the π-AB effect and AB effect are rather robust against the temperatures up to 22 K (Figure 4c,d). The observation of π-AB effect near Dirac point confirms the spin-helical nature of topological surface states (that is Fermi arcs), which has been demonstrated by previous ARPES results.[86, 209, 210] Such topologically-protected Fermi arc transport can be more clearly revealed in non-local measurement geometry, in which the bulk influence is greatly suppressed and the surface-related contribution is amplified. Altshuler-Aronov-Spivak (AAS) effect has been observed in a much thicker nanowire, where the quantum confinement along circumference is missing and thereby surface sub-band related-AB effect is strongly suppressed.[90]

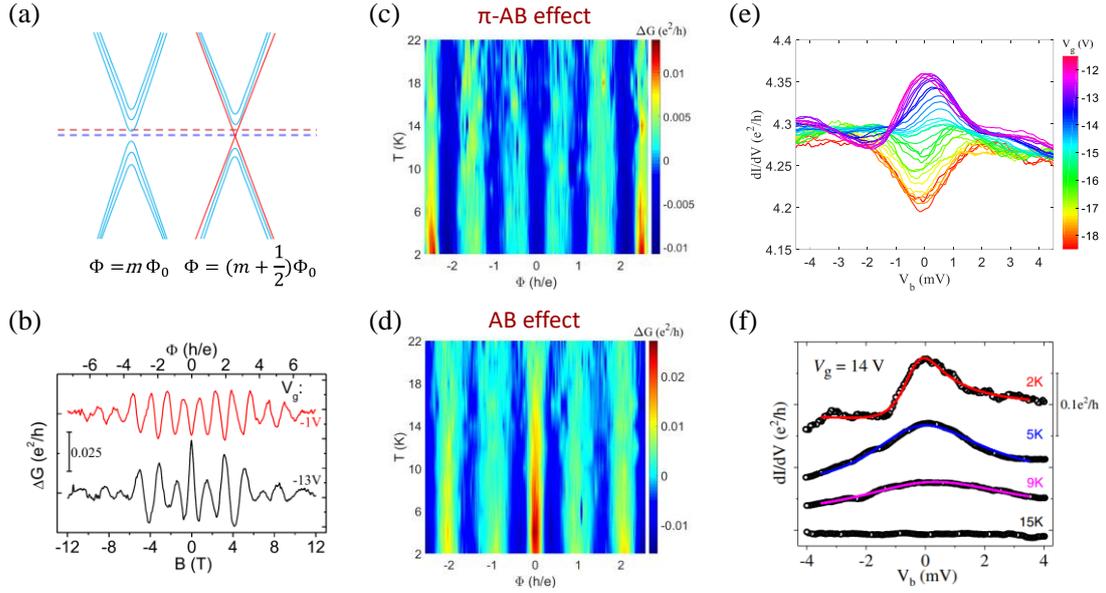

**Figure 4. AB effect and Fano interference.** (a) A schematic diagram of the surface energy bands when applying magnetic fields. (b) The AB oscillations when the Fermi level is located near (red curve) and away from the Dirac point (black curve), which correspond to π-AB and AB effect, respectively. (c,d) The temperature dependence of π-AB effect and AB effect. (e) The dI/dV spectra as a function of bias voltage $V_b$ for different gate voltages at 2 K. (f) The evolution of dI/dV spectrum with increasing temperatures. The solid lines correspond to the Fano fitting result. Panels b-d reprinted with permission from ref 91. Copyright 2017 American Physical Society. Panels e and f reprinted with permission from ref 102. Copyright 2018 American Physical Society.

**Fano Interference between Bulk and Surface States.** Fano has proposed that the interference between two scattering amplitudes, one due to scattering within a continuum of states and the second due to an excitation of a discrete state, would lead to the Fano resonance.[212] The Fano effect has been observed in many systems, including photonics,[213, 214] quantum dot,[215] carbon nanotubes,[216] single-electron transistor[217] and individual magnetic atoms.[218] The discussion about the AB effect above has shown the discrete surface sub-bands in thin topological semimetals nanowires due to the quantum confinement along the circumference. For the $Cd_3As_2$ nanowires with both discrete surface sub-bands and continuous bulk states, the Fano interference will also emerge.[102] The Fano effect will lead to a typical asymmetric profile about the differential conductance dI/dV spectrum (Figure 4e).[215, 218] Through tuning the gate voltage, a transition from zero bias peak (ZBP) to zero bias dip (ZBD) is observed (Figure 4e).[102] The ZBP occurs because the Fermi energy is located in the sub-band, while the ZBD occurs because the Fermi energy is located in the gap. Further measurements show that the ZBP gradually splits into two conductance peaks with increasing magnetic field. The linear splitting indicates the splitting is naturally related with Zeeman effect, giving the effective g factor ~32, which is consistent with pervious experiments in $Cd_3As_2$.[87] The asymmetry of the ZBP curve gradually disappears with increasing temperature, which is attributed to the thermal broadening of the sub-bands (Figure 4f). It is worth noting that the Fano effect can be observed at a relatively high temperature, which shows a large energy gap between sub-bands again, consistent with the AB effect.[90, 91] The Fano effect reveals a type of interaction between the bulk and surface states of topological semimetals, which will bring a phase shift into the Weyl orbit and a modification on the Weyl orbit related quantum oscillation frequency.[92, 95, 219]

**Quantum Hall effect.** Quantum Hall effect is featured with quantized Hall conductivity plateaus and simultaneously vanishing longitudinal resistance under magnetic fields. For conventional 2D electron system or trivial surface states, the Hall plateau is quantized as $gne^2/h$ with the degeneracy factor $g$ and integer $n$.[1, 220] For Dirac systems, the quantum Hall plateaus appear at $g(n + 1/2)e^2/h$, where the 1/2 term stems from the Berry phase $\pi$.[221] For the topological insulators, the degeneracy comes from the two parallel surfaces, and thus the filling factor $\upsilon = \left(n_t + \frac{1}{2}\right) + \left(n_b + \frac{1}{2}\right) = n_t + n_b + 1$, where $n_t$ ($n_b$) denotes the Landau index of top (bottom) surface.[222] Such 1/2 term in Hall conductance is the hallmark of topological surface states, which offers an effective tool to distinguish from trivial surface states.

For the Dirac semimetal $Cd_3As_2$, the two superposed Fermi arcs can form a closed Fermi loop due to chirality mixing effect (Figure 5a).[92] Such Fermi loop enables carriers to complete a cyclotron motion (indicated by black arrows in Figure 5a) on a single surface, which can give rise to the quantum Hall effect. Lin *et al.* systematically studied the quantum Hall effect in $Cd_3As_2$ nanoplate by tuning gate voltage and tilting the magnetic field direction (Figure 5b).[101] They observed quantized Hall plateaus at even filling factors as a magnetic field along the [112] crystal direction (Figure 5c). When rotating the magnetic field toward the [001] direction, namely $C_4$ rotation symmetry axis, the filling factors gradually evolve from $\upsilon = 2, 4, 6$ to $\upsilon = 3, 5$ (Figure 5d). They interpret it in the regime of topological phase transition in $Cd_3As_2$.[101] According to quantized Hall conductance $g(n + 1/2)e^2/h$, where 1/2 manifests the spin-helical nature of surface states, the odd filling factor $\upsilon = g(n + 1/2)$ reveals the topology of corresponding surface states (here consider degeneracy $g = 2$).[221, 222] When the magnetic field is along [001] direction, the $C_4$ symmetry is sustained, in which nontrivial Fermi-loop (formed by two Fermi arcs) surface states contribute to the odd-integer quantum Hall effect (Figure 5e).[11, 223] However, when the field is applied along [112] direction, the $C_4$ crystal symmetry is broken and the original Dirac nodes are gapped (Figure 5f).[78, 87, 224] Such broken symmetry would reconstruct the two Fermi arcs,[92] rendering the trivial surface states and leading to even-integer quantum Hall effect.

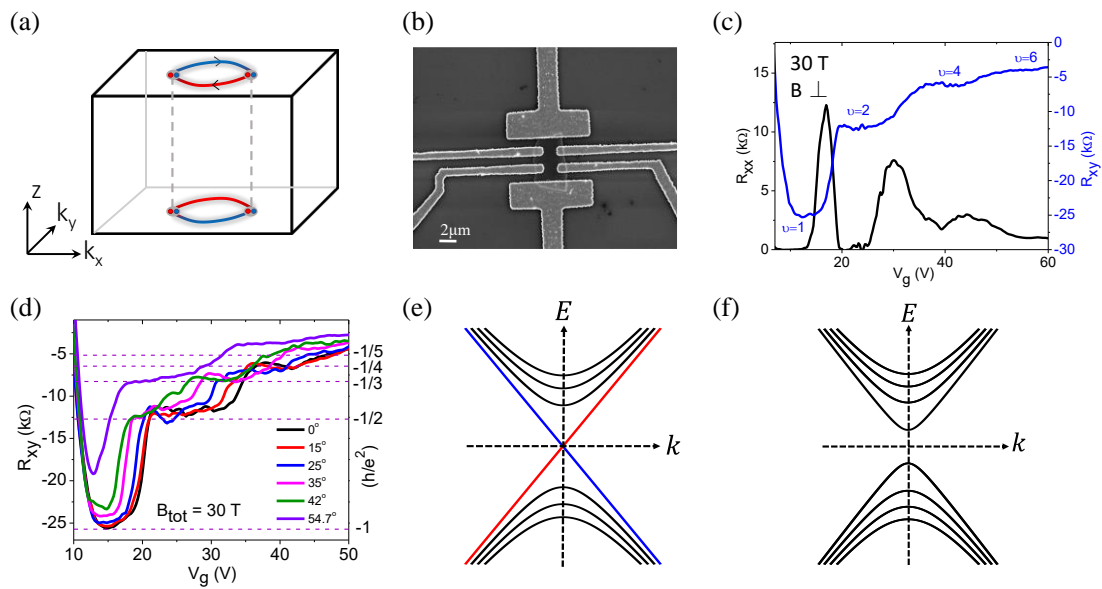

**Figure 5. Quantum Hall effect in 3D Dirac semimetals.** (a) Illustration of Fermi loops in a Dirac semimetal. On the surface, each Fermi loop is composed of two open Fermi arcs (red and blue denote the Fermi arcs with opposite chirality). (b) A SEM image of a $Cd_3As_2$ nanoplate Hall bar device. (c) Quantized Hall plateaus under a perpendicular magnetic field (along [112] direction). (d) Hall plateaus under a rotated magnetic field from 0° (along the [112] direction) to 54.7° (along the [001] crystal direction). (e) LLs under magnetic fields with $C_4$ rotational symmetry preserved. The lowest LL holds linear dispersion. (f) LLs under magnetic fields with $C_4$ rotational symmetry broken, which shows a gap-opening near the Dirac point. Panels b-f reprinted with permission from ref 101. Copyright 2019 American Physical Society.

Several other groups also report the observation of quantum Hall effect in thin Cd$_3$As$_2$ films and platelets.[97-100] The quantum Hall plateaus in Cd$_3$As$_2$ are found to be a series of even filling factors $v = 1, 2, 4, 6\ldots$, where $v = 1$ is explained by the splitting of the lowest Landau level.[97, 98] Nishihaya *et al.* attributed the observed quantum Hall effect to the quantum confinement induced bulk sub-bands in the thin film with thickness ~35 nm.[97, 99] The quantum Hall plateaus at even filling factors are ascribed to the confinement gap near the Dirac point. Besides the surface Fermi loop scenario and bulk sub-band regime, the inter-surface tunneling mediated by Weyl orbits is believed to be a route to result in the quantum Hall effect in 3D topological semimetals.[219] For Dirac semimetal Cd$_3$As$_2$, it remains a question whether the Weyl orbit can be preserved when the bulk Weyl nodes are gapped by magnetic fields.[87, 101] Galletti *et al.* report the variation of carrier density in Cd$_3$As$_2$ can lead to a notable phase shift of the quantum Hall effect.[225]

Until now, the underlying mechanism of quantum Hall effect observed in 3D Dirac semimetal Cd$_3$As$_2$ remains under debate. Several interpretations have been put forward as noted before, including topological surface states,[101, 226] quantum confinement induced bulk sub-bands[97, 99] and the Weyl orbits.[100] Recently, the quantum Hall effect has also been observed in a bulk ZrTe$_5$ sample, where the strong electron-electron interactions and induced periodic potential drive a 3D electronic system to quantum Hall regime under magnetic fields.[227-229]

**THE ELECTRONIC DEVICES**

Electronic properties of bulk and surface states render topological semimetals promising for device applications, such as, the linear energy dispersion for high-speed eletronics,[6-8] the spin-polarized Fermi arcs for magnetic memory devices,[86, 90, 91] and the topological nature of surface electrons for fault-tolerant quantum compuation.[107-109] Device applications that are being explored mainly include ultrafast broadband photodetectors based on broad-range and high-sensitive photo response of topological semimetals,[230-238] low-consumption spintronic devices that utilize the spin-polarized surface Fermi arc states[7, 209, 210, 239] and topological quantum computation that manipulates Majorana modes from the combination of helical nature and superconductivity.[107, 109, 203] Besides, the controlled topological phase transitions in Weyl/Dirac semimetals provide possibility for building topological spin (or charge) field effect transistors.[240-242] The discovered large thermoelectric response also makes topological semimetals promising for high-performance thermal detectors and energy converters.[156, 243-246]

**Ultrafast photodetectors.** The Dirac electronic systems,[247-253] including graphene, topological insulators, and Dirac semimetals are beneficial for infrared photodetection. Different from the 2D graphene and surface state of TI, the bulk states of 3D topological semimetals are more robust and not surface sensitive. Furthermore, the bulk structure renders the topological semimetals potentially have larger density of states, that can guarantee higher optical absorption and stronger light-matter interaction than those of atomically-thin graphene, which provides potential for high responsivity of photodetection.[231, 254, 255]

$Cd_3As_2$, a stable Dirac semimetal, has been identified to have many electronic

properties attractive for photodetections, including gapless bulk states,[78, 84] high carrier mobility[84, 87, 145] and ultrafast transient time.[230, 231] For example, the gapless band structure is promising for photodetection of low energy down to the terahertz frequency,[256-259] the high mobility for low-dissipation optoelectronic performance,[260, 261] and the ultrafast transient time for ultrafast photo response.[262, 263] Wang et al. reported the realization of an ultrafast broadband photodetector based on $Cd_3As_2$.[231] As depicted in Figure 6a, photocurrent signals can be recorded simultaneously when scanning the laser beam across the $Cd_3As_2$ nanoplate device. The laser beam is modulated with a mechanical chopper, and the short-circuit photocurrent signal is detected with a current preamplifier and a lock-in amplifier. The ultrafast transient relaxation time of $Cd_3As_2$ enables a very high response time of about 6.87 ps, which corresponds to a high frequency ~145 GHz (Figure 6b).[230, 231] Benefiting from the strong light interaction as bulk materials, the Dirac semimetal-based photodetector exhibits a high responsivity of 5.9 mA/W, which surpasses the performances of prototype graphene and many other 2D materials.[260, 264] Besides, the photodetector demonstrates a broadband wavelength response from 532 nm to 10.6 μm (Figure 6c), potential for a next-generation photodetector with a range extendable to far-infrared and terahertz.[257-259] Recently, an enhanced performance of wideband photodetector has been reported on the $Cd_3As_2$ heterojunction at room temperature.[232] The ultrafast broadband photodetector on Dirac semimetal provides excellent opportunities as a wonderful platform to realize various photonic applications, including high-performance optical communications, remote sensing, interconnects, surveillance and

spectroscopy.[231, 265, 266]

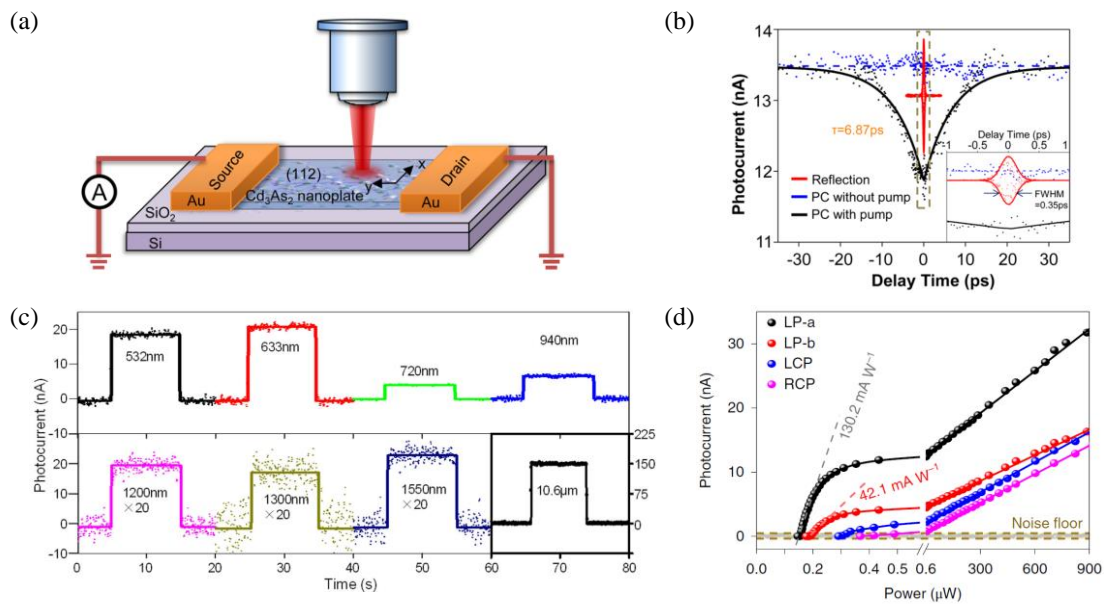

**Figure 6. Topological semimetal-based photodetectors.** (a) Schematic of scanning photocurrent measurement of $Cd_3As_2$ nanoplate devices. (b) Probe induced photocurrent with pump on (black) and off (blue) as a function of pump-probe pulse delay. A response time of 6.87 ps is deduced from the exponential fitting of the black solid line (see ref 231 for details). (c) Photocurrent responses of a $Cd_3As_2$ nanoplate device under different photoexcitation wavelengths (as marked). (d) Excitation power dependence of photocurrent of a $TaIrTe_4$ device at 4 μm excitation linear polarization (LP-a, LP-b) and circular polarization (LCP, RCP) excitations, respectively. A large photoresponsivity of 130.2 mA/W is accessed from the slope of linear fitting (grey dashed lines). Panels a-c reprinted with permission from ref 231. Copyright 2017 American Chemical Society. Panel d reprinted by permission from Springer Nature Customer Service Centre GmbH: Springer Nature Materials ref 238, copyright 2019.

The Weyl semimetal-based photodetectors, such as TaAs,[233] $WTe_2$,[234, 235] $MoTe_2$[236]

and TaIrTe$_4$,[237] have also been investigated. However, all these efforts are still limited to the experimental research of laboratories, and semimetal devices for practical applications are still missing. Recently, Ma *et al.* have realized a practical photodetection on type-II Weyl semimetal TaIrTe$_4$.[238] They studied the excitation power dependence of photocurrent of a TaIrTe$_4$ device at linearly polarized along *a* axis (LP-a), *b* axis (LP-b), left circular polarization (LCP) and right circular polarization (RCP) excitations (Figure 6d). TaIrTe$_4$ photodetector exhibits a large photoresponsivity of 130.2 mA/W with excitation wavelength 4 μm at room temperature, solving the long-lasting responsivity problem of semimetal-based photodetectors from the fundamental physical principle level (Figure 6d). Such large photoresponsivity enables semimetals for high-sensitive photodetectors with practically applicable values.[238] Besides, they have also observed an enhanced shift current, arising from Berry curvature singularity near the Weyl nodes.[267-270] This finding will stimulate more researches to enhance photo response through utilizing and manipulating the Weyl nodes.

**Spintronic devices.** Spintronic devices based on quantum materials have recently attracted great attention, including 2D materials, Rashba interfaces, topological insulator surface states and non-collinear antiferromagnets.[271-273] For example, the detection of spin transport has been extensively investigated in the topological insulators. Topological insulators are featured with insulating bulk states and gapless surface states with spin momentum locking.[3, 4] When passing a dc current (that is applying an electric field $E$), the Fermi surface shifts, and a net momentum $k_e$ is acquired, producing a net spin-polarized current with polarization $S$ due to the spin-helical nature (Figure 7a).[274] Generally, such spin current can be detected by ferromagnetic (FM) tunneling contacts, where the measured voltage is dependent on the relative orientation between the FM magnetization M and the induced spin polarization S.[275] Such electrical detection of current induced spin polarization has been demonstrated in many TI materials, including $Bi_2Se_3$,[274, 276] $Bi_{1.5}Sb_{0.5}Te_{1.7}Se_{1.3}$,[277] $(Bi_{1-x}Sb_x)_2Te_3$,[278-280] $Bi_2Te_2Se$[281] and $BiSbTeSe_2$.[282] Furthermore, the spin-momentum locking property is very useful for magnetization switching *via* spin-orbit torque. It has been found the net spin accumulation in the topological insulator surface can effectively switch and manipulate the magnetization of the adjacent magnetic layer.[283-290] Besides the Dirac surface states of topological insulators, topologically trivial two-dimensional surface states with a large Rashba splitting can also lead to spin polarization in the presence of an electric current. Arising from the surface band bending effect, such Rashba state could result in an opposite sign of current-induced spin polarization to the case of topological insulator surface states.[275] Different from the Dirac surface states of TIs, the two-dimensional Rashba states possess a parabolic energy dispersion. Moreover, the Rashba state strongly depends on the interfacial electric field, and its spin helicity can be electrically tuned to opposite direction,[282] in contrast with the topological surface states.

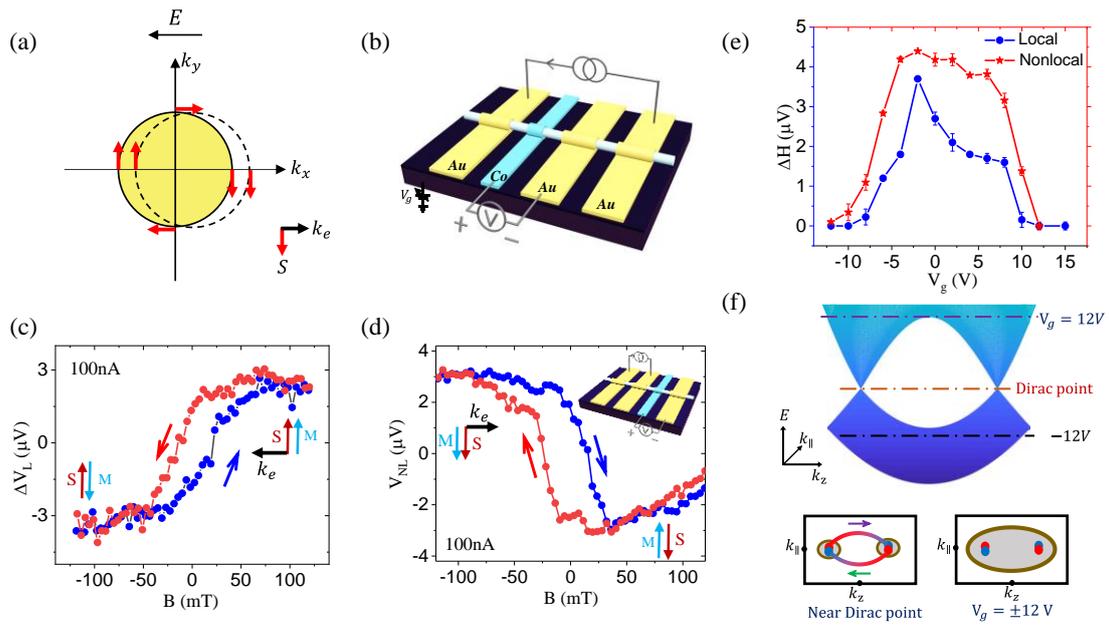

**Figure 7. Fermi arc spin transport in the Dirac semimetal Cd$_3$As$_2$ nanowire.** (a) Electric-current-induced spin polarization in the topological surface states. (b) Schematic of local spin detection measurements based on a Cd$_3$As$_2$ nanowire device. (c) Magnetic hysteretic loops measured with a dc current of 100 nA in the local regime. (d) Nonlocal spin detection results with a dc current of 100 nA. Inset denotes the non-local measurement geometry. (e) Gate dependence of window height ΔH of the magnetic hysteretic loop in the local and nonlocal measurements. ΔH represents the strength of spin signals. (f) Mechanism of topological phase transition in Dirac semimetal Cd$_3$As$_2$. Top: bulk energy band. Bottom: surface Fermi arcs. Each Fermi arc connects a pair of bulk Weyl Fermions with opposite chirality (blue and red balls). Shaded grey regions represent the Fermi energy pocket. Panel a reprinted by permission from Springer Nature Customer Service Centre GmbH: Springer Nature Nanotechnology ref 274, copyright 2014. Panels b-f reprinted with permission from ref 240. Copyright 2020 American Physical Society.

The Fermi arcs of topological semimetals are suggested to have spin momentum locking property, that is, the spin is in-plane locked to the momentum.[7, 78, 209, 210] Such spin momentum locking property is similar to the surface states of topological insulators, promising for current-induced spin polarization and realistic spintronic applications. Figure 7b shows the local transport measurement geometry of the spin polarized surface states of a Dirac semimetal $Cd_3As_2$ nanowire.[240] The dc current is applied between the outermost two gold electrodes, and voltage signal is measured between the inner magnetic (Co) and gold electrodes. The magnetic field is applied along the Co magnetization direction, in-plane perpendicular to the nanowire axis. Due to the spin-helical nature of topological surface states, a charge current would induce a spin-polarized current. When a bias current +100 nA is applied, a net in-plane spin **S** is established (Figure 7c). As the spin **S** is parallel to the magnetization **M**, there appears a high voltage. Otherwise, a low voltage emerges. Therefore, sweeping the in-plane magnetic field would lead to a hysteretic loop near the $B = 0$ T, as seen in Figure 7c. To clarity the spin signal of surface states, the background from bulk states has been subtracted in local measurement regime. When the bias current is reversed (-100 nA), the spin orientation of surface states would be reversed due to spin momentum locking, thus resulting a reversed hysteretic loop. The spin transport from topological surface states can also be detected in a non-local measurement geometry (inset in Figure 7d), where the dc current is applied to the two neighboring gold electrodes and the voltage is measured between the nonlocal Co and gold electrodes. Compared to the local regime, the non-local measurements can greatly reduce bulk contribution and pick out the surface spin signal. As shown in Figure 7d, a hysteretic loop is observed, indicating that the surface states are topologically protected and robust against the disorders and defects. Similar spin signals of surface Fermi arc states are also detected in Weyl semimetal $WTe_2$.[239]

In contrast to the case of TIs, the topological surface states of $Cd_3As_2$ can be changed to trivial states through Lifshitz transition.[78, 87, 165] When the Fermi level is located near the Dirac point, the surface projection of Dirac node is a point and the Fermi arcs become manifest (Figure 7f). As tuning the Fermi level away from the Dirac point, the density of Fermi arc states would be decreased. Particularly, when the Fermi level is further tuned beyond the Lifshitz transition point ($V_g = \pm 12V$), then projections of Dirac nodes would merge into an ellipsoid, rendering trivial surface states (Figure 7f). Such gate-triggered Lifshitz transition contributes to a gate-tunable surface spin signal, where the spin signal reaches the peak near Dirac point and gradually disappears as away from the Dirac point, demonstrating a large switch on/off ratio with electric field (Figure 7e). Similar topological phase transition is also reported, for example, in few-layer phosphorene, the electric field can convert the initial trivial insulating phase to the topological Dirac semimetal state, rendering the off-to-on state of charge and spin current.[241, 242] These electrical-induced topological phase transitions realize the switch between topological and trivial phase, and thus gives rise to the spin (charge) on/off effect, providing the possibility of establishing a multifunctional topological field effect transistor that can manipulate spin and charge transport simultaneously.

Despite of the potential outlook, the experimental evidence of spin transport on topological semimetals still remains scarce up to date. Further research is needed to demonstrate the spin transport from surface Fermi arcs. Besides, much attention should be devoted to the corresponding spintronics research, such as spin-orbit torque (SOT)-induced magnetization switching,[283-290] spin pumping[291, 292] and spin injection.[293]

**Majorana zero energy mode.** Topological superconductors are an exotic form of matter that host Majorana fermions at the boundaries.[294-296] Majorana zero energy modes have raised much concern due to their potential use as topological qubits to perform fault-tolerant quantum computation.[297, 298] Since it's hard to trace intrinsic topological superconductors in nature, several schemes have been proposed to artificially create topological superconductors. One method to produce topological superconductivity is combining the inherent electron-hole symmetry in a *s*-wave superconductor with the helical nature of electron states in quantum materials.[299] Signatures of Majorana mode have been observed in superconducting hybrid nanostructures of semiconductors with strong Rashba spin-orbit coupling,[300-303] topological insulators[304, 305] and ferromagnetic chains.[306]

With gapless bulk states and nontrivial helical arc-like surface states, topological semimetals are predicted to possess both intrinsic and extrinsic (proximity effect-induced) topological superconductivity.[107, 223, 307, 308] Unconventional superconductivity has been observed in topological semimetals, utilizing some specific techniques, such as high-pressure,[309, 310] point contacts,[311, 312] *etc.*

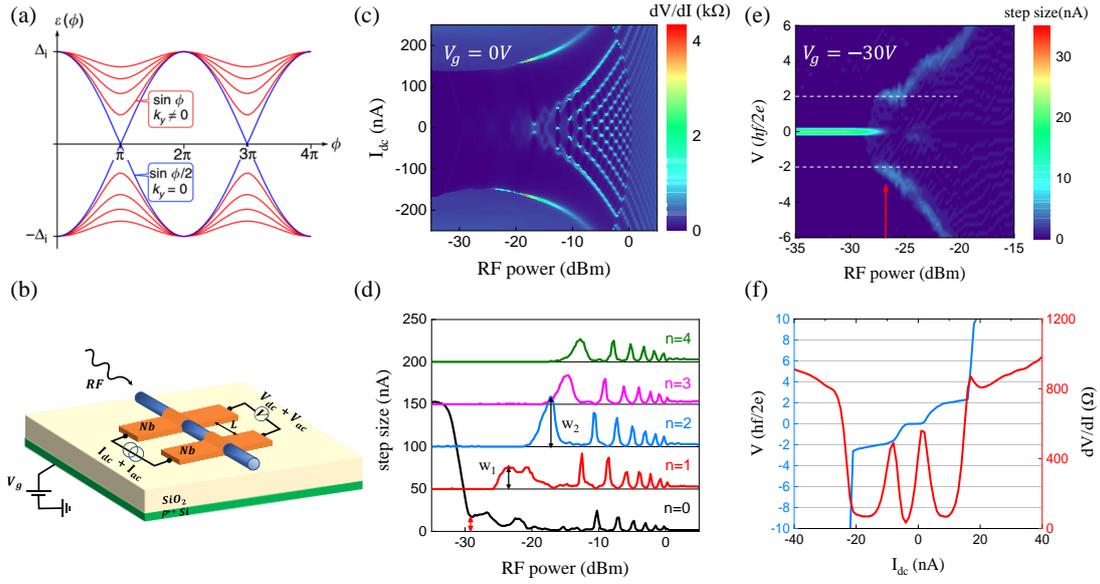

**Figure 8. 4π-periodic supercurrent in a Dirac semimetal.** (a) Typical Andreev spectrum of a topological surface-based Josephson junction. The gapless 4π-periodic Andreev bound states are denoted by blue lines, while gapped modes in red denote the conventional 2π-periodic Andreev case. (b) Schematic of the RF measurement setup. (c) Map of differential resistance dV/dI *versus* dc current bias $I_{dc}$ and RF power at $V_g$ = 0 V. (d) The Shapiro step size for different integer indices with varying RF power. The red arrow denotes the residual supercurrent of n = 0 step. (e) Map of the Shapiro step size *versus* normalized voltage V and RF power for $V_g$ = -30 V. (f) The $I_{dc}$ dependent V and dV/dI at $V_g$ = -30 V under RF power = −26.75 dBm, indicated by the red arrow in (e). Panel a reprinted with permission under a Creative Commons CC BY License from ref 304. Copyright 2016 Springer Nature. Panels b-f reprinted with permission from ref 109. Copyright 2018 American Physical Society.

Notably, the combination of s-wave superconductivity with helical surface states can produce gapless 4π-periodic Andreev bound states, where the bound states cross at zero energy (a Majorana mode).[304, 313-316]. As depicted by Figure 8a, the peculiarity of gapless Andreev bound states is its 4π-periodicity with superconducting phase difference ϕ across the junction. Such 4π-periodicity is a predominant signature of topological superconductivity and Majorana mode. In a Josephson junction geometry, the fusion of two nearby Majorana modes would produce an ordinary fermion (charge-e), instead of Cooper pairs (charge-2e), modifying the periodicity of current-phase-relation from 2π to 4π.[301] The experimental verification of 4π-periodic supercurrent generally relies on radio-frequency (RF) irradiation to produce Shapiro steps in a Josephson junction.[304] In RF measurements, the Josephson junction was irradiated *via* a coaxial line, in which an RF driving current is coupled with dc current bias $I_{dc}$ together to induce phase-locked Shapiro steps (Figure 8b). For a pure 4π-periodic supercurrent, only even Shapiro steps would appear. Experimentally, the suppression of n = 1 step is much greater than other odd steps due to capacitive effect or Joule overheating.[317, 318] Therefore, the existence of 4π-periodic supercurrent always accompanies the suppression (even missing) of n = 1 Shapiro step. Moreover, theoretical calculations indicate that a residual supercurrent of first node at n = 0 step, is also a direct and compelling evidence of 4π-periodic contribution.[304, 318]

Figure 8b shows the schematic of the RF measurement setup based on a Nb-Cd$_3$As$_2$ nanowire-Nb Josephson junction. A dc bias signal $I_{dc}$ is superimposed on the $I_{ac}$ and concurrently electrical signals are acquired in the pseudo four-probe current-voltage geometry. When $V_g$ = 0 V, the evolution of the Shapiro steps is distinguished from the conventional pattern of 2π-periodic supercurrent (Figure 8c). Both the n = 1 step suppression and residual supercurrents are clearly observed (Figure 8d), together suggesting the existence of 4π-periodic supercurrent. As tuning the gate voltage $V_g$ to enhance the surface state contribution ($V_g$ = -30 V), the odd (n = 1) Shapiro step is completely diminished (Figure 8e,f). Besides the 4π-periodic superconductivity, the proximitized surface Fermi arcs could give rise to supercurrent oscillations under in-plane magnetic fields, as lately reported in hybrid Nb-$Cd_3As_2$ nanoplate Josephson devices.[319] Such gate and field modulation of Fermi arc-related superconductivity in Dirac semimetal $Cd_3As_2$ provides possibilities to manipulate superconducting quantum states, which are potential building blocks for topological quantum computation. Notably, the missing n=1 step is also reported in exfoliated $Cd_3As_2$ under low frequency irradiation.[108] People have also observed the 4π-periodic supercurrent in Dirac semimetal $Bi_{1-x}Sb_x$.[320] However, the 4π-periodic contribution in $Bi_{1-x}Sb_x$ arises from the bulk Dirac cones instead of surface Fermi arcs. As for the different origins of 4π-periodic supercurrent, maybe it comes from the intrinsic discrepancy of the two materials themselves.

Besides the proximity effect above, another route to induce superconductivity in topological materials is through intercalation or doping.[8] Superconductivity has been observed in metal-intercalated topological insulators, such as Cu-intercalated, Sr-intercalated $Bi_2Se_3$ and $Bi_2Te_3$ and In-doped SnTe of a proper stoichiometry.[321-326] The Weyl semimetal $WTe_2$ and $MoTe_2$ have also been reported to exhibit superconducting behaviors under pressure or doping.[310, 327-329] As for the question of whether such superconductivity is topological, it remains to be further investigated.

# CONCLUSION AND OUTLOOK

We have discussed the abundant transport properties of topological semimetals, from the perspectives of bulk Weyl/Dirac fermions and surface Fermi-arc states. We also demonstrate the breakthroughs on the topological semimetal-based electronic devices, including the ultrafast broadband photodetectors, spin topological field effect transistors and Majorana zero modes toward topological qubits. Despite the prominent and extraordinary advances in topological semimetal researches, there are still numerous issues waiting to be tackled. In the following, we will discuss several important aspects and look toward the future.

**Controllable synthesis.** To fully exploit the nontrivial properties, high-quality sample is the most significant ingredient. However, controlling the crystal quality and morphology of topological nanostructures has remained a difficult issue. Nowadays, the most common synthesis methods of nanostructures are chemical vapor deposition (CVD) *via* vapor-liquid-solid (VLS) mechanism. Many previous experiments have revealed the complexity of the growth method, the control of which requires full understanding of crystal nucleation and growth mechanism.[330-333] Besides, numerous environmental variants also need to be carefully considered, including the local growth temperature, substrate types, gaseous precursors and growth pressure. The lack of *in situ* diagnostic tool further adds to the difficulty in controlling nanostructure growth. To solve these growth problems, we need to acquire a better knowledge of the growth mechanics, through *in situ* sample characterization, such as transmission electron microscope (TEM) growth studies. Meanwhile, theoretical simulations can be made to aid the synthesis process. Beyond the CVD and VLS methods, the development of molecular beam epitaxy (MBE) provides possibilities for the controllable growth of nanostructures. To obtain more desirable transport properties, the search for more semimetal materials is equally significant. Recent theoretical works predict thousands of topological materials based on symmetry knowledge,[334-336] which gives us an excellent guidance to the search of practical materials.

**Manipulation of quantum states.** With nontrivial Fermi-arc surface states and Dirac/Weyl bulk states, topological semimetal nanostructures have demonstrated a series of quantum states, such as discrete surface states, dissipationless edge states, spin-polarized Fermi arcs and Majorana bound states. For real-life device applications, the control and manipulation of quantum states are of outmost importance. Electrical modulation provides an alternative route to perform manipulations. Under the electric fields, the quantum states can be switched from nontrivial to trivial status and *vice versa*. For example, the surface Fermi arcs of Dirac semimetals can be changed to trivial Fermi loops under gate modulation.[110, 240] While in few-layer phosphorene, the electric field can transform the initial trivial insulating phase to the topological Dirac semimetal state.[241, 242] These electrical-induced topological phase transitions realize the switch between topological and trivial phase, and thus gives rise to the spin (charge) on/off effect, providing possibilities for establishing topological field effect transistors.

Applying an external magnetic field provides another route to manipulate the quantum states. The magnetic flux can modulate the accumulated phase within the trajectory process, realizing the manipulation of surface states, such as oscillatory superconductivity of proximitized Fermi arc surface states.[319] The magnetic modulation of Fermi arc-related superconductivity is helpful to manipulate superconducting quantum states, which are building blocks for topological quantum computation.[319] Besides, controlling device geometry size also provides a method to manipulate quantum states, especially boundary states with different dimensions.[130]

**Spin-based field effect transistors and memories.** Characterized by spin-helical Fermi arc surface states, topological semimetals have become next-generation candidates for spintronic applications. The helical nature of Fermi arc has been revealed by spin-ARPES measurements,[209, 210] and the transport measurements give a direct evidence of spin polarization.[240] Apart from the surface Fermi arcs, topological semimetals also possess highly conductive bulk states, which makes the surface spin detection difficult. In the future, electrical detection of Fermi-arc-related spin current is expected to be tried on more topological semimetal nanostructures, where bulk contribution is suppressed and surface Fermi arc topology is amplified. The spin current can be used as information transistors and logic devices.[337-340] For example, the spin-based field effect transistor can provide on and off operations *via* the manipulation of spin current. Considering the evolution of Fermi arc topology under Lifshitz transition, field-effect on spin signal is possible in Dirac semimetals.[78, 87, 165] Besides, the spin current can also be viewed as the transfer of angular momentum, which is useful for magnetization switching *via* spin-orbit torque, promising for magnetic random accessory memory. Furthermore, benefiting from the exotic topology of Fermi arcs, the spin current tends to be robust against environmental perturbance, which helps to

increase the stability of future spintronic devices.[240]

**Topological quantum computation.** Topological quantum computation presents another exciting frontier direction. Topological quantum computation is an approach to fault-tolerant quantum computation, based on the braiding of non-Abelian anyons. Majorana zero modes (also Majorana modes) are believed as the simplest realization of non-Abelian anyons.[341] Well-separated Majorana modes obey non-Abelian braiding statistics, providing a natural implementation of a topological computer. With helical surface states, the topological semimetal nanowire provides a promising platform to realize Majorana modes when combining with s-wave superconductor. For braiding operations, the formation and location of Majorana modes must be well controlled. For a nanowire, the control of Majorana modes can be realized through electrostatic gating or magnetic flux.[342-345] The application of topological computer requires a system not only to demonstrate braiding but also to read out the initial and final state (fermion parity). One of the most practical read-out schemes is to combine Majorana modes with transmon qubits, as depicted in Figure 9a.[345] The setup is composed of two blocks of Cooper pair boxes (as Figure 9b) embedded in a transmission line resonator that forms a transmon qubit. In the T junction, the three overlapping Majorana modes can effectively form a single Majorana mode, resulting in a total of six Majorana modes in the system $\gamma_A$, $\gamma_B$, $\gamma_C$, $\gamma_D$, $\gamma_E$, $\gamma_F$ (Figure 9a). Coulomb coupling between Majorana modes can be tuned through magnetic flux. Figure 9c demonstrates a typical braiding process, where the three steps together realize the exchange between Majorana 1 and 2. Before and after the braiding operations, the fermion parity of each island is read out by the transmon qubit.

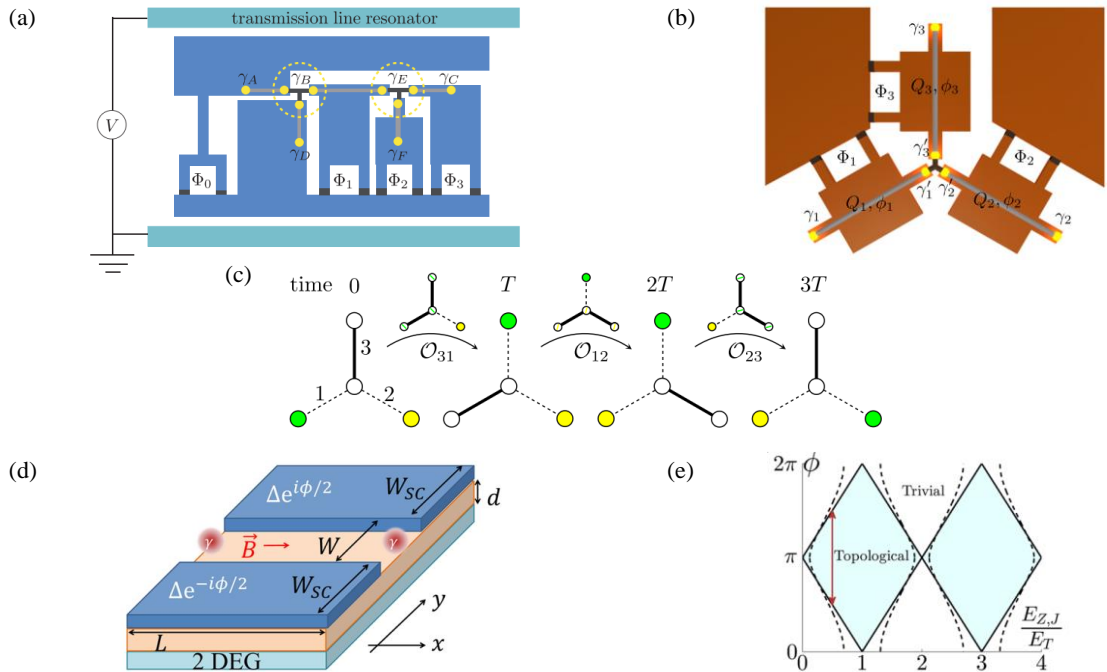

**Figure 9. Toward the Majorana braiding in one-dimensional and two-dimensional architecture.** (a) Flux-controlled demonstration of non-Abelian Majorana statistics. Each superconducting island (blue color) contains a nanowire with two Majorana modes at its ends (yellow circles). (b) Three Cooper pair boxes connected at a T-junction. (c) Schematic of the three steps of braiding operation at a T-junction. The white circles denote the Majorana modes with large Coulomb splitting and the color ones represent the Majorana modes with small splitting. Insets above each arrow denotes the intermediate state between neighboring steps. (d) A planar Josephson junction based on 2DEG with strong Rashba spin-orbit interaction and two superconductors with phase difference $\phi$. (e) Phase diagram as a function of Zeeman energy $E_{Z,J}$ and phase difference $\phi$. The solid and dashed lines represent the phase boundaries of a junction with perfect and non-perfect interface, respectively. Panel a reprinted with permission from ref 345. Copyright 2013 American Physical Society.

Panels b and c reprinted with permission under a Creative Commons CC BY-NC-SA License from ref 344. Copyright 2012 IOP Publishing and Deutsche Physikalische Gesellschaft. Panels d and e reprinted with permission under a Creative Commons Attribution 4.0 International License from ref 349. Copyright 2017 American Physical Society.

Despite the growing number of experimental breakthroughs,[300, 303, 346-348] scalable networks of Majorana qubits are still challenging to realize in the one-dimensional nanowire systems due to the practical difficulty in delicate parameter tuning and geometric implementation. To overcome the challenges, an alternative platform of two-dimensional electron gas (2DEG) system, which confines Majorana channels within planar Josephson junction, has been put forward (Figure 9d).[349] Under an in-plane Zeeman field, the junction enters the topological superconducting state and supports two Majorana modes at its end points (red balls in Figure 9d), similar to one-dimensional proximitized nanowire case. Remarkably, phase difference offers an additional tuning parameter to control over the topological superconducting phase which is not explored in one-dimensional systems.[350, 351] The 2DEG-based planar Josephson junctions can sustain topological superconductivity over a much larger parameter space (light blue regions in Figure 9e), addressing the cumbersome issue of fine tuning. Based on the advances of planar Josephson junctions, Zhou *et al.* have proposed a topological X-shaped junction to braid and fuse multiple pairs of Majorana modes, providing a versatile platform to probe non-Abelian statistics in the future.[352]

To conclude, long-standing developments have been achieved in the topological semimetal researches, from physical properties to electronic devices. We believe that the advance and maturity of topological semimetals will inspire an insight into modern industry and provide possibilities for largescale applications.

## ACKNOWLEDGMENTS

This work was supported by National Natural Science Foundation of China (No. 91964201, No. 61825401 and No. 11774004), and National Key Research and Development Program of China (No. 2018YFA0703703 and No. 2016YFA0300802).

## VOCABULARY

**nanostructures,** a class of structures that have at least one dimension between 1 to 100nm and usually possess size-dependent physical properties due to quantum-confinement effects; **Fermi arcs,** discontinuous segments of a two-dimensional Fermi contour, which are terminated onto the projections of the Weyl fermion nodes on the surface; **topological electronics (topotronics),** the study and application of the topological materials and related electron transport properties in solid-state devices; **photodetector,** the device to detect light or other electromagnetic radiation, usually based on the external or internal photoelectric effect; **spintronics,** a portmanteau meaning spin transport electronics, focusing on the study of the intrinsic spin of the electron and its associated transport phenomena in solid-state devices; **quantum computation,** a type of computation that makes uses of quantum-mechanical phenomena such as superposition and entanglement, to perform operations on data.